\documentclass[aps,prb,10pt,twocolumn,amsmath,amssymb,floatfix]{revtex4-2}
\usepackage[english]{babel}
\usepackage{graphicx,bbm}
\usepackage[utf8]{inputenc}
\usepackage[colorlinks,linkcolor=blue,citecolor=blue,urlcolor=blue]{hyperref}

\begin{document} 
\title{Doping $S=1$ antiferromagnet in one-dimension}
\author{J. Prokopczyk}
\author{J. Herbrych}
\affiliation{Institute of Theoretical Physics, Faculty of Fundamental Problems of Technology, Wroc{\l}aw University of Science and Technology, 50-370 Wroc{\l}aw, Poland}
\date{\today}
\begin{abstract}
Antiferromagnetic ground states, when doped, give rise to rich and complex phenomena, prompting detailed investigations in various spin systems. Here, we study the effect of doping on the one-dimensional $S = 1$ antiferromagnetic Heisenberg model (AFM). Specifically, we investigate how the presence of holes affects the static and dynamic (frequency-dependent) spin-spin correlations of the two-orbital Hubbard-Kanamori chain. The latter, at half-filling and in the strong-interaction limit, maps onto an $S = 1$ Heisenberg model. For moderate interactions, an orbital resonating-valence-bond (orbital-RVB) state emerges up to doping levels of $x \lesssim 0.3$. A detailed analysis of interaction strength $U$ and doping concentration $x$ reveals that this phase inherits the key features of spin excitations found in the half-filled case --- namely, a gapped spin spectrum and ``coherent'' magnon behavior up to a wavevector $q$ determined by the Fermi vector, $2k_\mathrm{F} = \pi(1 - x)$. Furthermore, our results uncover an additional broad, incoherent spectral weight for $q \gtrsim 2k_\mathrm{F}$ at high frequencies. Finally, we show that near the transition to a ferromagnetic phase, a previously unidentified spiral-like state emerges, characterized by spin excitations reminiscent of the $J_1$-$J_2$ Heisenberg model.
\end{abstract}
\maketitle

\section{Introduction}

The doping of antiferromagnetic (AFM) ground states remains a central challenge in condensed matter physics. Since the discovery of high-temperature superconductivity in doped cuprates nearly five decades ago, both electron and hole doping of AFM systems have been shown to give rise to a variety of novel quantum phases and exotic excitations \cite{Lee2006,Armitage2010,Keimer2015,Fradkin2015,Baltz2018}. This has motivated extensive experimental and theoretical investigations, the latter particularly focused on understanding the properties of the $S=1/2$ Heisenberg model, the doped single-orbital Hubbard model, and its effective low-energy description - the $t$-$J$ model. Owing to the versatility of materials \cite{Lake2005,Lake2013,Scheie2022,Scheie2025} and cold-atom quantum simulators \cite{Mazurenko2017,Hilker2017,Sompet2022,Richaud2022,Bourgund2025}, these systems can also be realized in reduced dimensions (e.g., chains and ladders), enabling direct comparisons with state-of-the-art numerical methods capable of solving the corresponding Hamiltonians exactly.

Less is known about the doping of $S=1$ AFM systems \cite{Xu2000,Ammon2000,Malvezzi2001,Laurell2024}. Such scenarios are relevant for iron-based superconductors and other materials with multiple active orbitals close to the Fermi surface \cite{Zorko2002,Baskaran2008,Glasbrenner2015,Wang2015,Fabbris2017,Gao2024}. In such compounds, the on-site ferromagnetic Hund's interaction forces the spins of the electrons on different orbitals to be aligned in parallel \cite{Georges2013}. Consequently, in the limit of large interactions, the low-energy spin excitations are composed of $S>1/2$ moments. A prominent example of such a scenario is AFM $S=1$ Heisenberg compounds in one-dimension (1D) \cite{Zaliznyak2001,Kenzelmann2002,Maximova2020,Nag2022,Jelinek2023}, which exhibit a famous Haldane gap.

In a 1D AFM system of $S=1/2$ fermions, introducing charge carriers can significantly alter the magnetic properties, giving rise to incommensurate spin correlations \cite{Parschke2019}, spin-charge separation \cite{Ogata1990,Hilker2017}, and even superconducting tendencies under certain conditions. Understanding how doping affects $S=1$ AFM chains is especially intriguing due to the presence of the Haldane gap and the topological nature of the undoped ground state \cite{Haldane1983,Pollmann2010,Becker2017}. Canonically, the ground-state of the $S=1$ Heisenberg model is explained via the connection to the Affleck-Kennedy-Lieb-Tasaki (AKLT) valence bond (VB) state with only nearest-neighbor spin-spin correlations \cite{Affleck1987}. The isotropic $S=1$ Heisenberg model is a gapped state with exponentially decaying correlations without any spin long-range order [akin to the resonating valence bond (RVB) state]. In principle, introducing holes (or electrons) into the RVB state encapsulates Phil Anderson's idea \cite{Anderson1987,Liang1988} for the formation of Cooper pairs and, eventually, the superconducting state in cuprates \cite{Baskaran1987}. 

It was recently shown \cite{Mierzejewski2024} that the RVB-nature of $S=1$ ground state survives till quite large doping at interaction strength comparable to the kinetic energy and, indeed, leads to the superconducting tendencies \cite{Patel2017,Patel2020,Laurell2024}. Such a phase (at finite doping) was coined the orbital-RVB state. In this work, we will focus on a detailed analysis of the magnetic properties of the latter state. We achieved this by investigating the static and dynamic (frequency-dependent) spin-spin correlations (i.e., the dynamical spin structure factor) within the two-orbital Hubbard-Kanamori model. The Schrieffer-Wolff transformation of the latter in the large interaction limit (and at half-filling) yields an $S=1$ Heisenberg model. Our findings indicate that the spin excitations crucially depend on the interaction strength and doping $x$. Within the orbital-RVB phase, the low-frequency spin spectrum resembles that of an $S=1$ antiferromagnet (with "coherent" magnon mode and a spin gap). However, at large frequencies, an additional feature appears as a broad, incoherent spectrum at short wavelengths. Finally, we also discuss the spiral-like state that emerges at the transition between orbital-RVB and ferromagnetic phases.

The paper is structured as follows. In Section~\ref{sec:model}, we introduce the two-orbital Hubbard-Kanamori model together with the main quantities and the methods used in this work. In Section~\ref{sec:phasediagram}, we describe how the $S=1$ ground state emerges in the two-orbital model and discuss the general properties of the doped $S=1$ state. Section~\ref{sec:mag} presents the main results of our work: the analysis of the static (Sec.~\ref{sec:magstat}) and dynamic (Sec.~\ref{sec:magdyn}) spin structure factors. The conclusions are given in Section~\ref{sec:conc}.

\section{Model \& Methods}
\label{sec:model}

In this work, we will focus on the two-orbital Hubbard-Kanamori (HK) model \cite{Dagotto2011,Georges2013,Georges2024} in one spatial dimension:
\begin{eqnarray}
H_\mathrm{HK}&=& \sum_{\gamma\gamma^\prime\ell\sigma} t_{\gamma\gamma^\prime}
\left(c^{\dagger}_{\gamma\ell\sigma}c^{\phantom{\dagger}}_{\gamma^\prime\ell+1\sigma}+\mathrm{H.c.}\right)+\Delta_\mathrm{CF}\sum_\ell n_{1\ell}\nonumber\\
&+& U\sum_{\gamma\ell}n_{\gamma\ell\uparrow}n_{\gamma\ell\downarrow}
+U^\prime \sum_{\ell} n_{0\ell} n_{1\ell}\nonumber\\
&-& 2J_\mathrm{H} \sum_{\ell} \mathbf{s}_{0\ell} \cdot \mathbf{s}_{1\ell}
+J_\mathrm{H} \sum_{\ell} \left(P^{\dagger}_{0\ell}P^{\phantom{\dagger}}_{1\ell}+\mathrm{H.c.}\right)\,.
\label{eq:ham2o}
\end{eqnarray}
Here $c^\dagger_{\gamma\ell\sigma}$ ($c^{\phantom{\dagger}}_{\gamma\ell\sigma}$) represent electron creation (annihilation) operator at orbital $\gamma=\{0,1\}$, site $\ell=\{1,\dots,L\}$, and with spin projection $\sigma=\{\uparrow,\downarrow\}$. $\Delta_\mathrm{CF}$ stands for crystal field splitting, $n_{\gamma\ell}=n_{\gamma\ell\uparrow}+n_{\gamma\ell\downarrow}$ is density of the electrons, $U$ ($U^\prime=U-5/2J_\mathrm{H}$) is intra-orbital (inter-orbital) Hubbard interaction, $J_\mathrm{H}$ is the Hund exchange coupling, $\mathbf{s}_{\gamma\ell}=(1/2)\sum_{a,b}c^\dagger_{\gamma\ell a}\mathbf{\sigma}_{ab}\, c^{\phantom{\dagger}}_{\gamma\ell b}$ is the spin operator (with $\mathbf{\sigma}_{ab}$ as the Pauli spin matrices), and $P^{\dagger}_{\gamma\ell}=c^\dagger_{\gamma\ell\uparrow}c^\dagger_{\gamma\ell\downarrow}$ represent pair hopping operator. In the following, we will use a noninteracting kinetic energy span $W=4t$ as a unit of energy. We will consider an open-boundary system at fixed magnetization $S^z_\mathrm{tot}=\sum_{\ell,\gamma}s_{\gamma,\ell}=0$ and various hole doping levels $x=1-n$, where $n=\sum_{\gamma\ell} n_{\gamma\ell}/L=N/2L$ (with $N$ as the total number of electrons and $L$ as the system size). Since both Hubbard and Hund interactions originate from Coulomb many-body forces, we will fix the Hund coupling to $J_\mathrm{H}/U=0.25$ throughout the work. Past analysis of the HK model suggests that varying the $J_\mathrm{H}/U$ ratio does not significantly alter the conclusions \cite{Patel2020,Jazdzewska2023}. Nevertheless, for completeness, at the end of Sec.~\ref{sec:magdyn} we present results for various $J_\mathrm{H}/U$ ratio.

The main quantities of interest in this work are static and dynamical spin structure factors (SSF) at zero temperature. The static SSF is given by
\begin{equation}
S(q)=\frac{1}{L}\sum_{\ell,m} \mathrm{e}^{\imath q(\ell-m)}\,\langle \mathbf{S}_\ell\,\mathbf{S}_m \rangle\,,
\end{equation}
with $\mathbf{S}_{\ell}=\mathbf{s}_{0\ell}+\mathbf{s}_{1\ell}$ and $\langle \cdot\rangle=\langle \mathrm{gs}|\cdot|\mathrm{gs}\rangle$ as the expectation value in the ground-state. Note that in the limit of large interaction, $U\,,J_\mathrm{H}\gg t$, the local magnetic moments are given by $\mathbf{s}^2_{\gamma\ell}=3/4$ while $\mathbf{S}^2_\ell=2$. 

The dynamical SSF is defined as
\begin{equation}
S(q,\omega)=\frac{1}{L}\sum_{\ell} \mathrm{e}^{\imath (\ell-L/2) q}\,
\langle\langle \mathbf{S}_{\ell}\,\mathbf{S}_{L/2} \rangle\rangle_{\omega}\,.
\label{eq:sqw}
\end{equation}
Here
\begin{equation*}
\langle\langle A_{\ell}\,B_{m} \rangle\rangle_{\omega}=-\frac{1}{\pi}\mathrm{Im}\langle \mathrm{gs}|A_{\ell}\frac{1}{\omega^{+}-(H-\epsilon_\mathrm{GS})}B_{m}|\mathrm{gs}\rangle
\end{equation*}
represent the imaginary part of Green's functions with $\omega^{+}=\omega+i\eta$ and $\eta$ as an internal broadening (here always set to $\eta=2\Delta\omega$, where $\Delta\omega$ is a frequency resolution). $|\mathrm{gs}\rangle$ represents the ground-state wavevector with energy $\epsilon_\mathrm{GS}$. Note that, to reduce the computational resources, in Eq.~\eqref{eq:sqw} we perform only one out of two usual sums over the system size $L$ in the Fourier transform. If the system length $L$ is large enough for $\langle\langle A_{\ell}\,B_{c} \rangle\rangle_{\omega}$ correlations to decay to approximately zero, the analysis only from the system's center $c=L/2$ is sufficient \cite{Kuhner1999}. We explicitly checked that this is the case for all parameters considered in this work.

The ground-states of the Hamiltonians and the Green's functions examined in this study were obtained using the density matrix renormalization group (DMRG) method~\cite{White1992,Schollwock2005} within the single-site framework~\cite{White2005}. Dynamical correlation functions were computed via dynamical-DMRG~\cite{Jeckelmann2002,Nocera2016}, employing the correction-vector method and Krylov decomposition to evaluate spectral functions in frequency space directly~\cite{Nocera2016}. During the DMRG simulations, up to $M=2048$ states were retained, allowing us to assess the HK model~\eqref{eq:ham2o} for system sizes as large as $L=60$ sites, with truncation errors below $10^{-6}$. It is important to note that the artificial broadening $\eta$ used in the dynamical-DMRG method, which can be interpreted as the effects of disorder, finite temperature, or measurement-device resolution, hinders the magnon gap $\Delta_S$ from the spectrum. Detailed analysis of the latter was presented in Ref.~\cite{Jazdzewska2023,Mierzejewski2024}.  

\section{Doping $S=1$ antiferromagnet}
\label{sec:phasediagram}

\begin{figure}[!t]
\includegraphics[width=1.0\columnwidth]{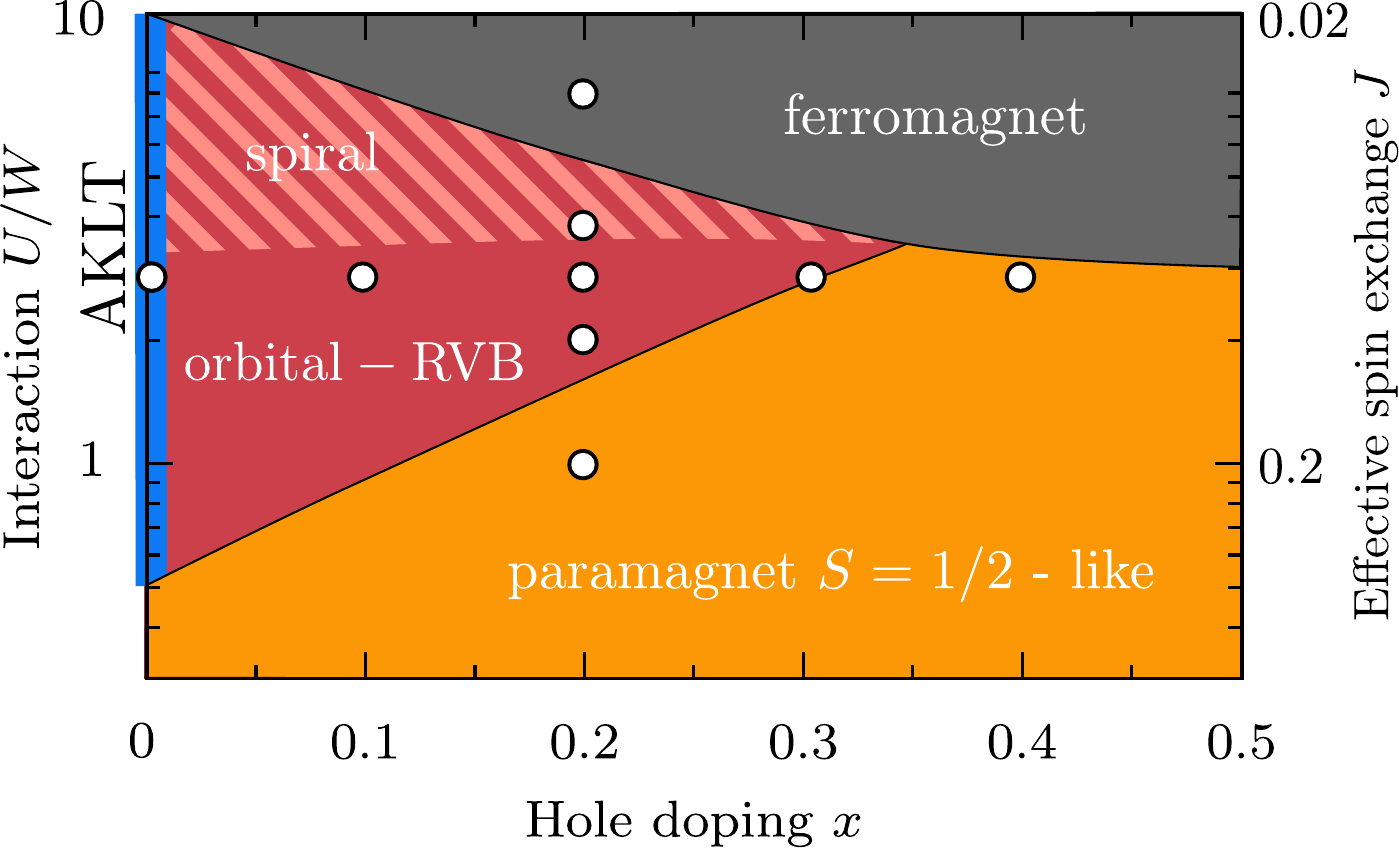}
\caption{Sketch of the interaction $U$-hole doping $x$ phase diagram of the two-orbital Hubbard-Kanamori model with degenerated bands. The left $y$-axis represents the interaction strength $U/W$, while the right $y$-axis represents the effective interaction strength $J=2t^2/(U+J_\mathrm{H})$ ($W=4t=2$). Note the log-scale. At $x=0$ and $U>U_\mathrm{c}$ the $S=1$ AKLT-like state is stabilized \cite{Jazdzewska2023} (equivalent to $S=1$ Heisenberg model in the $U\,,J_\mathrm{H}\gg t$ limit). The orbital-RVB state with nontrivial topological properties can be found for $0<x\lesssim 0.3$ and $U\sim 2W$~\cite{Mierzejewski2024}. The orbital-RVB region with possible spiral phase is estimated based on the results presented in Sec.~\ref{sec:magstat}. For $x\ne0$ and $U,J_\mathrm{H}\gg t$ ferromagnetic order is stabelized~\cite{Moreo2025}. A short-range paramagnet (similar to the doped single-band Hubbard model at small $U$) can be found for \mbox{$U\lesssim W$}. White points depict the $(U,x)$ value at which the dynamical spin structure factor $S(q,\omega)$ is evaluated in Sec.~\ref{sec:magdyn}.}
\label{fig:sketch}
\end{figure}

The HK model exhibits a rich magnetic phase diagram as a function of the system parameters. For $n=2$, the interaction-induced phase transition between the noninteracting result (for $U<U_\mathrm{c}\simeq W/2$) and AFM Haldane phase was presented in Ref.~\cite{Jazdzewska2023}. The low-energy (spin) excitations of the latter, in the limit of large interaction \mbox{$U\,,J_\mathrm{H}\gg t$}, can be described (through Schrieffer-Wolff transformation) by the $S=1$ Heisenberg model
\begin{equation}
H_\mathrm{S}=J\sum_\ell \mathbf{S}_{\ell}\cdot\mathbf{S}_{\ell+1}\,,
\label{eq:heis}
\end{equation}
with spin exchange $J=2t^2/(U+J_\mathrm{H})$. Here, the three projections of $S=1$ are built out of local (on-site) triplets of the atomic limit of \eqref{eq:ham2o}:
\begin{equation*}
|1\rangle=\left|
\begin{matrix}
\uparrow \\
\uparrow
\end{matrix}
\right\rangle\,,\quad
|-1\rangle=\left|
\begin{matrix}
\downarrow \\
\downarrow
\end{matrix}
\right\rangle\,,\quad
|0\rangle=
\frac{1}{\sqrt{2}}\left|
\begin{matrix}
\uparrow \\
\downarrow
\end{matrix}
\right\rangle+
\frac{1}{\sqrt{2}}\left|
\begin{matrix}
\downarrow \\
\uparrow
\end{matrix}
\right\rangle\,.
\end{equation*}
The studies of \eqref{eq:ham2o} indicated \cite{Jazdzewska2023} that such $S=1$ description is valid not only in the $U\to\infty$ limit, but already for rather low values of interaction $U>U_\mathrm{c}\simeq W/2$. The latter marks the interaction-induced phase transition between trivial and topological (akin to the AKLT state) ground-states. The transition is manifested by the appearance of, e.g., spin gap $\Delta_\mathrm{S}\ne0$, edge states, and finite string order parameter. On the other hand, for finite doping $x\ne 0$ and at $U,J_\mathrm{H}\gg t$, the double-exchange mechanism dominates and the system orders ferromagnetically \cite{Zener1951,deGennes1960,Anderson1955}. The analysis of the spin excitations in this region was recently shown in Ref.~\cite{Moreo2025}. See Fig.~\ref{fig:sketch} for sketch of the interaction-doping, $U$-$x$, phase diagram. 

In this work, we focus on the case of degenerated bands away from half-filling ($t_{00}=t_{11}=0.5$, $t_{01}=t_{10}=0$, $\Delta_\mathrm{CF}=0$, $W=2$, and $n\ne1$) in the most challenging region of interaction strength $U\sim{\cal O}(W)$. For such parameters, overall AFM-like ordering is expected \cite{Parschke2019}. Our previous investigation \cite{Mierzejewski2024} of doped $S=1$ state (within two-orbital Hubbard-Kanamori model) indicated that there exists a large region of $U$-$x$ parameters where the characteristic properties of the topologically nontrivial AKLT state survive (see Fig.~\ref{fig:sketch}). Such a state with AKLT properties at finite doping was coined as the orbital-RVB state \cite{Patel2020,Mierzejewski2024}. Interestingly, the topological properties of the latter coexist with superconducting tendencies \cite{Patel2017} as measured by binding energy and pair-pair correlations. 

It is worth noting that lifting band degeneracy (e.g., by $\Delta_\mathrm{CF}\ne0$ or $t_{11}\ne t_{00}$) may result in other exotic phases. In the case of orbital differentiation, one of the orbitals can undergo a Mott phase transition. At the same time, the other remains delocalized (metallic), i.e., the system can undergo a so-called orbitally-selective Mott transition (OSMP) \cite{Vojta2010,Georges2013}. Here, the competition between superexchange (AFM) and double-exchange (FM) mechanisms leads to block~\cite{Rincon2014,Rincon2014-2,Mourigal2015,Herbrych2018,Herbrych2019,Herbrych2020} and block-spiral~\cite{Herbrych2020-2} magnetic order. As we show below, remnants of some of these phases are also present in degenerated bands.

\begin{figure}[!t]
\includegraphics[width=1.0\columnwidth]{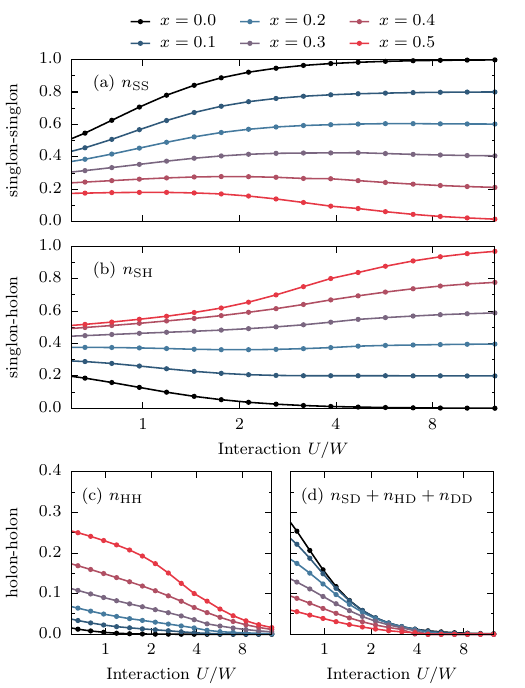}
\caption{Interaction $U/W$ dependence of the atomic-level states, i.e., average number of (a) site singlons $n_\mathrm{SS}$ (i.e., each orbital occupied by one electron); (b) orbital singlons $n_\mathrm{SH}$ (i.e., with one electron on one orbital and hole on another); (c) site holons $n_\mathrm{HH}$ (i.e., empty sites); and (d) contributions of sites with double occupancies $n_\mathrm{SD}+n_\mathrm{HD}+n_\mathrm{DD}$. See text for details.}
\label{fig:filling}
\end{figure}

Doping the system with holes, $n=1-x$, leads to the appearance of new states in the atomic limit:
\begin{equation*}
\left|
\begin{matrix}
\uparrow \\
0
\end{matrix}
\right\rangle\,,
\left|
\begin{matrix}
0 \\
\uparrow
\end{matrix}
\right\rangle\,,
\left|
\begin{matrix}
\downarrow \\
0
\end{matrix}
\right\rangle\,,
\left|
\begin{matrix}
0 \\
\downarrow
\end{matrix}
\right\rangle\,.
\end{equation*}
Note that the HK model \eqref{eq:ham2o} is particle-hole symmetric, i.e., our results are valid also for electron doping, $n=1+x$ (provided that one considers degenerated bands). The above states differ from those expected in the "standard" single-band Hubbard model. Here, each holon (the site where one of the orbitals is empty) carries a $\uparrow,\downarrow$-spin flavor. Consequently, while the doping of the single-orbital Hubbard model introduces spinless holons (or doublons) in the $S=1/2$ AFM background (encapsulated in the $t$-$J$ model), for the two-orbital Hubbard model, doping introduces $S=1/2$-like objects in the $S=1$ AFM background. Note also that at the atomic level, an empty site (both orbitals empty) is possible, provided the overall desired $n$ is achieved. However, our analysis indicates that such states have vanishing weight in the ground state. In Fig.~\ref{fig:filling}, we show the doping dependence of the density of various atomic-level states. We use the standard definition of (here orbital resolved) singlons and holons, i.e., \mbox{$n_{\gamma \ell}^\mathrm{S}=n_{\gamma \ell \uparrow}+n_{\gamma \ell \downarrow}-2n_{\gamma \ell \uparrow}n_{\gamma \ell \downarrow}$} and \mbox{$n_{\gamma \ell}^\mathrm{H}=1-n_{\gamma \ell \uparrow}-n_{\gamma \ell \downarrow} + n_{\gamma \ell \uparrow}n_{\gamma \ell \downarrow}$}, respectively. Next, we define density of site singlons \mbox{$n_\mathrm{SS}=\sum_{\ell}n_{0\ell}^\mathrm{S}\,n_{1\ell}^\mathrm{S}/L$} (sites occupied by two electrons, each at one of the orbitals), orbital singlons \mbox{$n_\mathrm{SH}=\sum_{\ell}\left(n_{0\ell}^\mathrm{S}\,n_{1\ell}^\mathrm{H}+n_{1\ell}^\mathrm{S}\,n_{0\ell}^\mathrm{H}\right)/L$} (sites occupied by one electron at one of the orbitals), and site holons \mbox{$n_\mathrm{HH}=\sum_{\ell}n_{0\ell}^\mathrm{H}\,n_{1\ell}^\mathrm{H}/L$}. It is evident from the presented results that in the $U\gg t$ limit only site singlons and orbital singlons survive.

\section{Magnetic properties}
\label{sec:mag}

\begin{figure}[!t]
\includegraphics[width=1.0\columnwidth]{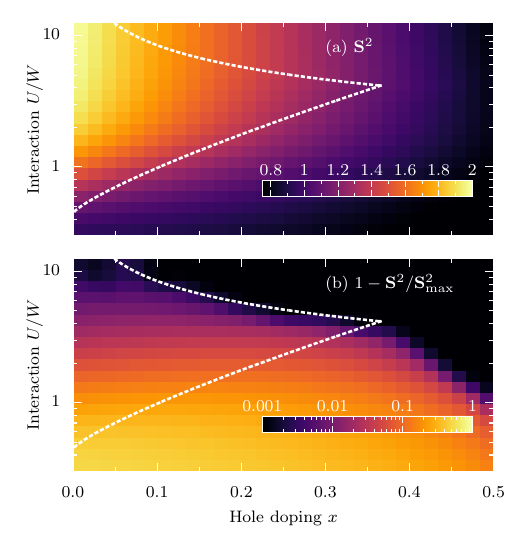}
\caption{(a) Average magnetic moment squared $\mathbf{S}^2=(1/L)\sum_i \mathbf{S}^2_i$ dependence on the hole doping $x$ and interaction $U$. The limiting ($U\gg t$) case $\mathbf{S}^2=2$ for $x=0.0$ ($\mathbf{S}^2=0.75$ for $x=0.5$) indicate an effective $S=1$ ($S=1/2$) magnetic moment. (b) The same data plotted as a fraction of the maximal possible magnetic moment \mbox{$\mathbf{S}^2_\mathrm{max}(x)=2\cdot(1-2x)+0.75\cdot 2x$}. The results are presented as $1-\mathbf{S}^2/\mathbf{S}^2_\mathrm{max}$ in the color log-scale. White dashed lines represent approximate borders of the region supporting the orbital-RVB state \cite{Mierzejewski2024}.}
\label{fig:moments}
\end{figure}

As discussed in the previous section, it was shown \cite{Jazdzewska2023,Mierzejewski2024} that the above \mbox{$S=1$} Heisenberg model description is valid not only in the $U\gg t$ limit, but also in the regions where the magnetic moments are not yet fully developed $\mathbf{S}^2=S(S+1)\ne2$. For $x=0$ this is true for $U>U_\mathrm{c}$, while for $x\ne0$ in the so-called orbital-RVB state, i.e., for $x\lesssim 0.3$ and $U\sim{\cal O}(W)$. Interestingly, our analysis of the average magnetic moment, $\mathbf{S}^2=(1/L)\sum_i \mathbf{S}^2_i$, indicates that in the orbital-RVB state the magnetic moments are never maximized to the $\mathbf{S}^2_\mathrm{max}(x)=2\cdot(1-2x)+0.75\cdot 2 x$ value. The data presented in Fig.~\ref{fig:moments} indicates that $\mathbf{S}^2=\mathbf{S}^2_\mathrm{max}(x)$ only for $x=0\,,U\gg t$ and when the system orders ferromagnetically for $x\ne0$. 

\subsection{Static structure factor $S(q)$}
\label{sec:magstat}

\begin{figure}[!t]
\includegraphics[width=1.0\columnwidth]{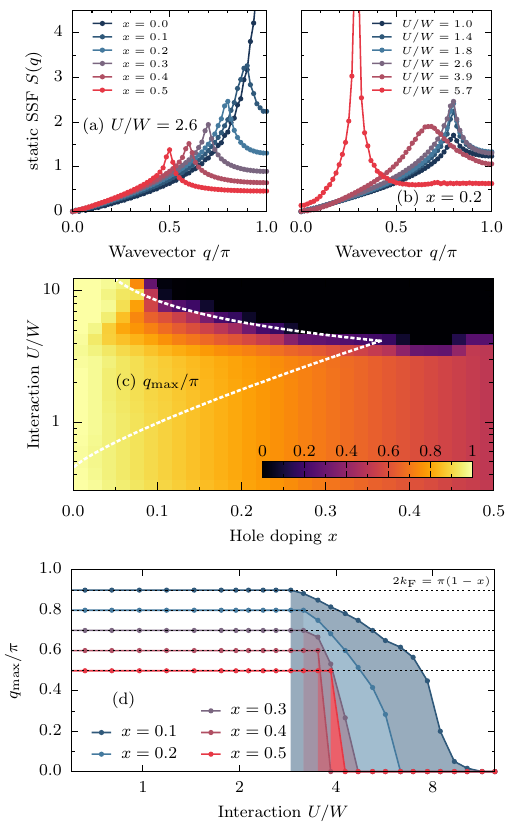}
\caption{(a,b) Static spin structure factor $S(q)$ evaluated for (a) fixed interaction strength $U/W=2.6$ and various hole dopings $x=0.0,\dots,0.5$; and for (b) fixed doping $x=0.2$ and various interaction strength $U/W=1.0,\dots,5.7$. (c) The position of maximum $q_\mathrm{max}$ of $S(q)$ in the $U$-$x$ plane of parameters. (d) Detailed interaction $U/W$ dependence (note the log scale) of $q_\mathrm{max}$ for $x=0.1,0.2,0.3,0.4,0.5$. Shaded region depict value of $U$ for which the spiral phase ($q_\mathrm{max}<2k_\mathrm{F}$) is stabilized.}
\label{fig:sq}
\end{figure}

Let us now focus on the magnetic properties of a doped $S=1$ antiferromagnet. In Fig.~\ref{fig:sq}, we present results for the static SSF $S(q)$. In panels (a) and (b), we present results for fixed interaction strength $U/W=2.6$ and fixed doping $x=0.2$, respectively. The remaining panels show the position of the maximum of $S(q)$, $q_\mathrm{max}$, in the whole $U$-$x$ plane parameter range. A few conclusions are evident from the presented results: within the region in where the orbital-RVB state is supported (as well in the limit of small interaction $U<U_\mathrm{c}$), the $q_\mathrm{max}$ follows the (noninteracting) Fermi vector, $q_\mathrm{max}=2k_\mathrm{F}=\pi(1-x)$. Such behavior is akin to the dependence of $S(q)$ of the single-band Hubbard model. Specifically, in the latter model (with the effective $S=1/2$ Heisenberg model for $x=0$ and $t$-$J$ model for $x\ne0$ in the $U/t\gg 1$ limit), doping does not destroy the overall AFM character of correlations, a phenomenon known as incommensurate antiferromagnetism. Our results indicate that similar behavior can be observed for the doped two-orbital Hubbard model for $U/W\lesssim 3$.

\begin{figure}[!t]
\includegraphics[width=1.04\columnwidth]{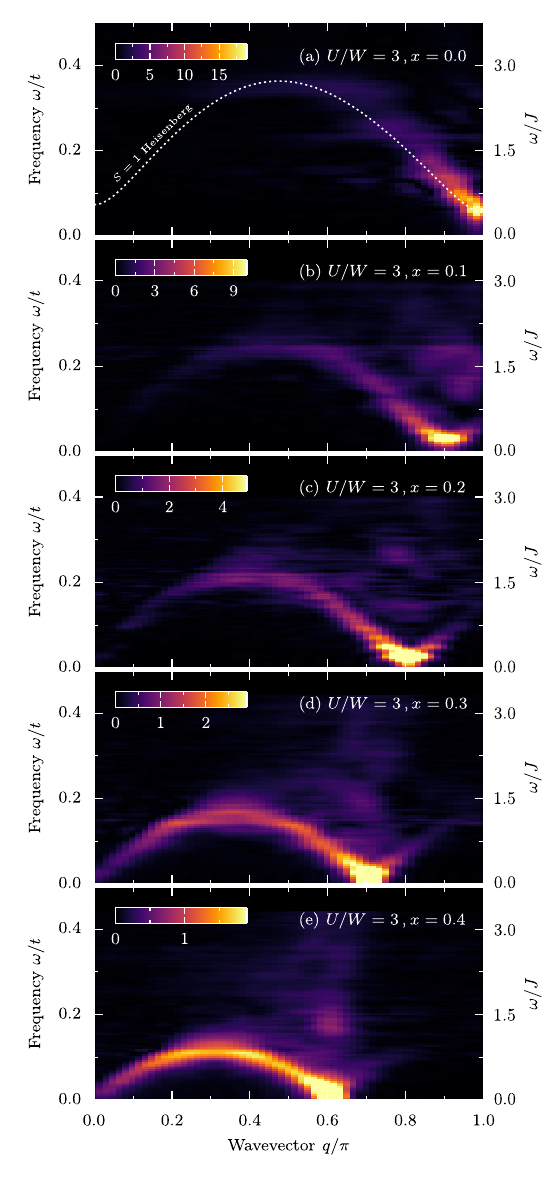}
\caption{Doping $x$ dependence of the dynamical spin structure factor $S(q,\omega)$ of the two-orbital Hubbard model for $U/W=3$. Panels (a-e) depict data for $x=0.0,0.1,\dots,0.4$, respectively. The white dashed line in panel (a) depicts the magnon dispersion relation of the $S=1$ Heisenberg model \cite{White2008}. In each plot, $100$ frequency points are shown (\mbox{$\Delta\omega=\omega_\mathrm{max}/100$}, where $\omega_\mathrm{max}$ is the maximal presented frequency). Left $y$-axis represents frequency $\omega$ in unit of hopping $t$, while right $y$-axis represents $\omega$ in units of $J=2t^2/(U+J_\mathrm{H})$.}
\label{fig:xdep}
\end{figure}

In contrast to $x=0$ - with one sharp phase transition between the trivial (paramagnetic) state and topological AKLT-like state at $U_\mathrm{c}$ - the finite doping $x\ne0$ exhibits two transitions \cite{Mierzejewski2024}. The lower one starts at $U_\mathrm{c}=W/2$ at $x=0$ and slowly increases till $U_\mathrm{c}(x)\sim 4W$ at $x\simeq0.35$. At large $U$ (and consequently large $J_\mathrm{H}$), the system undergoes a second, upper transition, entering the ferromagnetic state with trivial topological properties. Our analysis of static SSF indicates that the latter transition is marked by a continuous change of the position maximum of $S(q)$ from $q_\mathrm{max}=2k_\mathrm{F}$ (incommensurate AFM) to $q_\mathrm{max}=0$ (ferromagnet) [see Fig.~\ref{fig:sq}(b) and Fig.~\ref{fig:sq}(d)]. Such behavior is akin to the spiral phase (with $\langle \mathbf{S}_{i}\times\mathbf{S}_{i+1} \rangle\ne0$ instability) found in systems exhibiting OSMP, i.e., with lifted band degeneracy \cite{Herbrych2020-2}. Here, for example, for $x=0.1$, we find a wide range of interaction, $3\lesssim U\lesssim 10$, where such a state is stabilized. Note that the range of interaction values that support the spiral shrinks as the doping $x$ increases [see Fig.~\ref{fig:sq}(d) and sketch in Fig.~\ref{fig:sketch}].

\subsection{Dynamical structure factor $S(q,\omega)$}
\label{sec:magdyn}

\subsubsection{Doping dependence}

We start our analysis of the spin excitations [as measured by the dynamical SSF $S(q,\omega)$] with the $x=0$ result. In Fig.~\ref{fig:xdep}(a), we show magnon ($\Delta \mathbf{S}=1$) spectrum of the half-filled HK model \eqref{eq:ham2o} for $U/W=3$. As evident from the presented results, we achieve an excellent agreement with the expected $S=1$ Heisenberg model dynamics (shown as a dashed line \cite{White2008}). The dominant feature of the corresponding dynamical SSF is the gapped single-magnon branch (the Haldane gap $\Delta_S/J\simeq0.41$ at $q=\pi$) and vanishing intensity in the long-wavelength limit $q\to0$. Interestingly, for finite doping $ x\neq 0$ (provided that we are still within the orbital-RVB region), the main features of the spin spectrum remain unchanged, as shown in Fig.~\ref{fig:xdep}(b-c). Specifically: (i) all results within the orbital-RVB region exhibit a "coherent" magnon branch with the maximum at the $q=2k_\mathrm{F}=\pi(1-x)$ with finite (doping dependent) gap $\Delta_S(x)\ne0$. The internal broadening $\eta$ of Eq.~\eqref{eq:sqw} obstructs the detailed analysis of the doping dependence of the magnon gap $\Delta_S(x)$. We refer the interested reader to Refs.~\cite{Jazdzewska2023,Mierzejewski2024} for the analysis of the spin gap via the difference in ground-state energies of two magnetization sectors with different $S^z_\mathrm{tot}$. (ii) All results within orbital-RVB region, Fig.~\ref{fig:xdep}(a-c) $x< 0.3$, have vanishing intensity in $q\to0$ limit. On the other hand, the "coherent" magnon behavior at long wavelengths and closing of the spin gap is observed for $x\gtrsim0.3$, i.e., on the border of the orbital-RVB phase.

\begin{figure}[!htb]
\includegraphics[width=1.0\columnwidth]{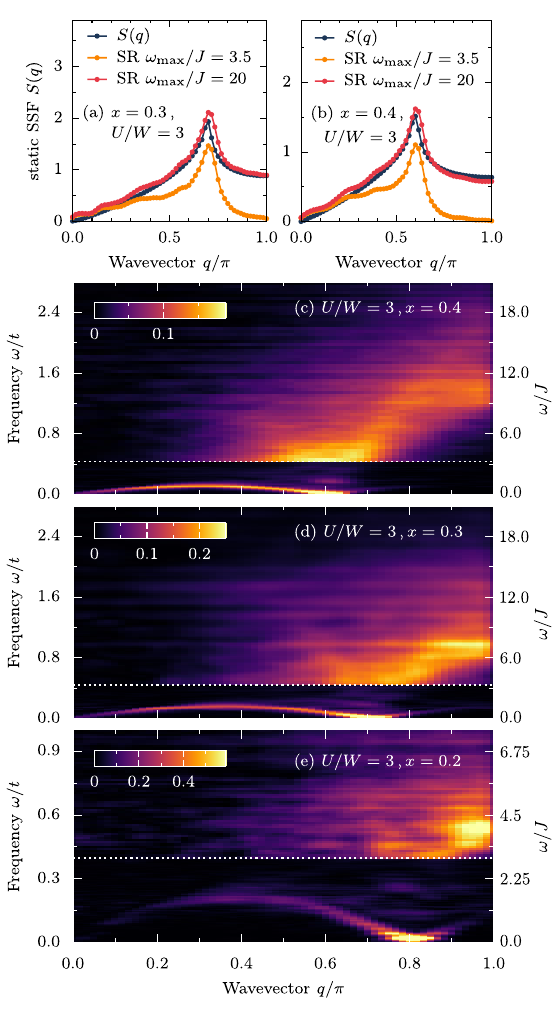}
\caption{(a,b) Comparison of the static structure factor $S(q)$ and the sum-rule (SR) of dynamical $S(q,\omega)$, Eq.~\eqref{eq:sr}, integrated up to frequencies $\omega_\mathrm{max}/J=3.5$ (the low-frequency dispersive branch of the spectrum) and $\omega_\mathrm{max}/J=20$. Calculated for $U/W=3$, (a) $x=0.3$ and (b) $x=0.4$. (c-e) High-frequency dynamical spin structure factor $S(q,\omega)$ evaluated for $U/W=3$, (c) $x=0.4$, (d) $x=0.3$, and (e) $x=0.2$. Here, the frequency resolution is set to $\Delta\omega/t=0.04$, $\Delta\omega/t=0.02$, and $\Delta\omega/t=0.01$, respectively. The results below the white dashed lines depict the one presented in Fig.~\ref{fig:xdep}(c-e).}
\label{fig:highom}
\end{figure}

Interestingly, results presented in Fig.~\ref{fig:xdep} clearly show that the spectrum of doped $S=1$ system exhibits strong suppression of the spectral intensity in the low-frequency region for $2k_\mathrm{F}<q<\pi$ wavevectors (or even lack of it for $x\gtrsim 0.2$). The missing weight is also evident by comparing the static $S(q)$ and the dynamic $S(q,\omega)$ SSF,  since the two are related by the sum-rule (SR)
\begin{equation}
S(q)=\frac{1}{2\pi}\int\limits_{0}^{\infty} \mathrm{d}\omega\,S(q,\omega)\,.
\label{eq:sr}
\end{equation}
For the half-filled case ($x=0$), the dynamical SSF presented in Fig.~\ref{fig:xdep}(a) fully saturates the above SR (not shown). On the other hand, the low-frequency spectral weight of $S(q,\omega)$ for $x\ne0$ [i.e., Eq.~\eqref{eq:sr} integrated up to $\omega_\mathrm{max}=3.5J$ in units of effective spin exchange $J=2t^2/(U+J_\mathrm{H})$] does not. Our results presented in Fig.~\ref{fig:highom}(a,b) indicate that the SR is fulfilled only for long-wavelengths $q\to0$, with missing weight for $q\simeq 2k_\mathrm{F}$ and lack of it for $q\gtrsim 2k_\mathrm{F}$. Consequently $S(q\gtrsim 2k_\mathrm{F},\omega)$ have to be finite for $\omega>3.5J$. Our analysis indicates that the "missing" spectral weight for $q\gtrsim2k_\mathrm{F}$ appears as very broad, very incoherent excitations in a large frequency region. For large doping, $x=0.4$, the latter reaches frequency $\omega_\mathrm{max}\simeq20J$. Decreasing doping systematically decreases $\omega_\mathrm{max}$. See Fig.~\ref{fig:highom}(c-e). Note also that the intensity of this mode is an order of magnitude smaller than the low-frequency magnons. Integrating Eq.~\eqref{eq:sr} up to $\omega_\mathrm{max}$ fully saturates to the expected value of $S(q)$ [see Fig.~\ref{fig:highom}(a,b)].

Such vanishing weight of short wavelengths \mbox{$q\to\pi$} spectrum and high-$\omega$ modes were already reported in the case of the spin excitations of the OSMP systems \cite{Herbrych2018,Herbrych2020} (i.e., with orbital differentiation). Also, experimental inelastic neutron scattering (INS) investigation of the low-dimensional OSMP compound BaFe$_2$Se$_3$ \cite{Mourigal2015,Monney2013} indicates on the presence of such modes. There, the high-frequency modes take the form of well-defined ("coherent"), momentum-independent excitations. It was shown \cite{Herbrych2020} that the latter depends on the value of the Hund exchange $J_\mathrm{H}$. Here, in the case of the doped $S=1$ Heisenberg model, the high-frequency mode is incoherent and dispersive. Nevertheless, the large frequency span $\sim 20J\sim J_\mathrm{H}$ could indicate that the Hund exchange also controls these high-$\omega$ features of the spectra.

\subsubsection{Hubbard interaction dependence}

\begin{figure}[!htb]
\includegraphics[width=1.0\columnwidth]{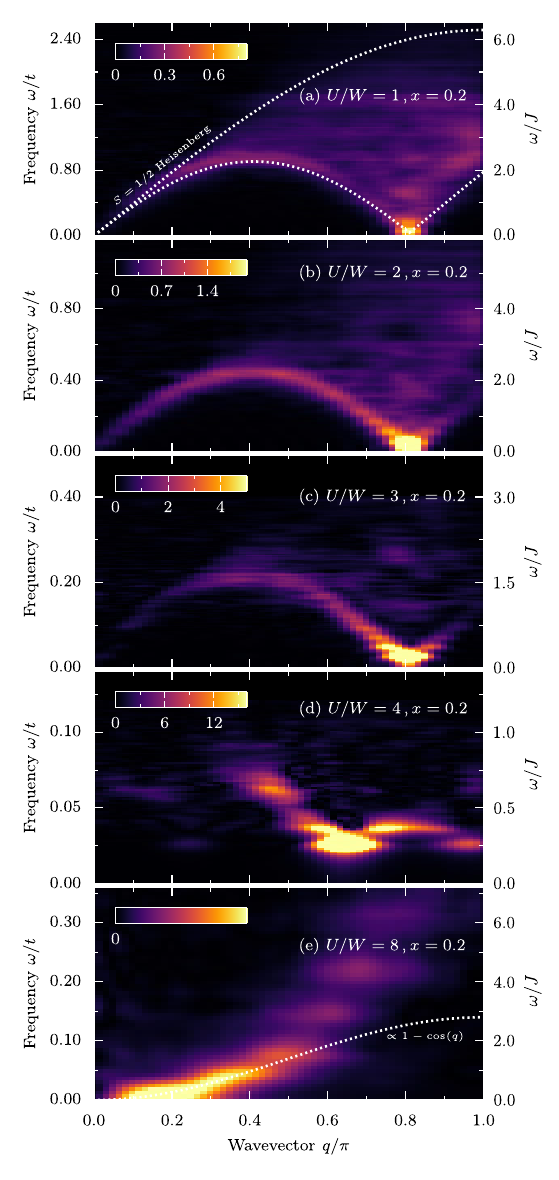}
\caption{Interaction $U$ dependence of the dynamical spin structure factor $S(q,\omega)$ for finite doping $x=0.2$. Panels (a-e) depict data for $U/W=1,2,3,4,8$, respectively. The white dashed line in panel (a) depicts the upper and lower bounds of the two-spinon continuum of the $t$-$J$ model at finite doping. The white dashed line in panel (e) depicts the standard magnon dispersion $\omega\propto 1-\cos(q)$ above ferromagnetically order state. In each plot, $100$ frequency points are shown ($\Delta\omega=\omega_\mathrm{max}/100$, where $\omega_\mathrm{max}$ is the maximal presented frequency). Left $y$-axis represents frequency $\omega$ in unit of hopping $t$, while right $y$-axis represents $\omega$ in units of $J=2t^2/(U+J_\mathrm{H})$.}
\label{fig:udep}
\end{figure}

In this section, comment on the interaction $U$ dependence of spin excitations for finite doping $x$. The results for half-filling ($x=0$) were presented in Ref.~\cite{Jazdzewska2023}. Here, in Fig.~\ref{fig:udep}, we focus on $x=0.2$.

At small interaction (below the orbital-RVB phase), the spin excitations resemble those of the single-orbital Hubbard model, i.e., the two-spinon continuum [see Fig.~\ref{fig:udep}(a) for $U/W=1$ result]. Such behavior is expected, as in the $U\to0$ limit the two-orbital HK model reduces to two noninteracting chains. In the latter case, the boundaries of the spin spectrum can be evaluated using the Lindhard formula \cite{Giamarchi2004}. In the case of single-orbital Hubbard mode, increasing the interaction strength leads to $S=1/2$ Heisenberg \cite{Laurell2022} or $t$-$J$ model \cite{Parschke2019} at $x=0$ or $x\ne0$, respectively. In the former case, a two-spinon continuum can be calculated with the help of the Bethe Ansatz \cite{Muller1981}. Both solutions, $U\to 0$ and $U\to\infty$, have the same functional form  [e.g., for $x=0$ and zero magnetization: $\omega_\mathrm{U}(k)=\alpha|\sin(k/2)|$ and $\omega_\mathrm{L}(k)=\alpha/2|\sin(k)|$], but with different energy scales $\alpha$, i.e., $\alpha=4t$ and $\alpha=\pi4t^2/U$, respectively. Note also that the latter value ($U\to\infty$ prediction) is not valid for finite $U$, especially for $U\sim W$ \cite{Laurell2022}. Furthermore, the two limits have different distributions of spectral weights with a maximum of the intensity at the top (bottom) of the continuum in the $U\to0$ ($U\to\infty$) limit, and as a consequence, increasing the interaction strength leads to a gradual transfer part of weight to the $\omega_\mathrm{L}$ line, preserving overall continuum structure \cite{Parschke2019,Laurell2022}.

Our $U/W=1$ results presented in Fig.~\ref{fig:udep} clearly show buildup intensity at the bottom of the continuum, with energy scale $\alpha\simeq2J$ [$J$ refers here to $S=1$ exchange $J=2t^2/(U+J_\mathrm{H})$]. However, in contrast to the single-orbital Hubbard model, with increasing $U$ the overall continuum is diminished with the emergence of a "coherent" magnon mode inside the orbital-RVB phase [see Fig.~\ref{fig:udep}(a-c)]. Note that around $U/W\simeq 2$ (for $x=0.2$) the spin (Haldane) gap opens \cite{Jazdzewska2023,Mierzejewski2024}; however, the finite broadening $\eta$ of our spectra prevents its detailed analysis.

An interesting aspect of our findings is that further increasing the interaction strength has a drastic effect on the spin excitations. Close to the transition between the orbital-RVB and ferromagnetic phase [$U/W=4$, Fig.~\ref{fig:udep}(d)], i.e., in the region of parameters where the maximum $q_\mathrm{max}$ of static SSF $S(q)$ does not follow the Fermi vector prediction, $q_\mathrm{max}<2k_\mathrm{F}$ [see Fig.~\ref{fig:sq}(d)]. Here, the spin spectrum resembles that of the $J_1$-$J_2$ Heisenberg model with AFM couplings $J_1\simeq J_2$ \cite{Ferrari2018}. The latter can give support to the spiral phase, in agreement with our static SSF investigation presented in Sec.~\ref{sec:magstat}. Note that a similar spiral phase was reported in the OSMP system, also as a precursor of the ferromagnetic phase \cite{Herbrych2019,Herbrych2020-2}.

Finally, for completeness, in Fig.~\ref{fig:udep}(e) we present results for the ferromagnetic phase (generically placed at $U\gg t$ and $x\ne0$; here considered at $U/W=8$ and $x=0.2$). In such a kinematically stabilized phase, electrons on orbital $\gamma$ minimize their kinetic energy when they can hop on the site where the electron on orbital $\gamma\prime$ has the same spin projection - thus leading to an overall ferromagnetic spin arrangement. Previous investigation of the systems with lifted orbital degeneracy (e.g., within OSMP phase) has shown \cite{Moreo2025} that the spin excitations above double-exchange ferromagnets follow the "standard" (Holstein–Primakoff-like) magnon dispersion, $\omega \propto 1-\cos(q)$, only in the limit of long-wavelengths $q\to0$. For short-wavelengths $q>\pi/2$, one observes a broad incoherent excitations due to scattering on the Stoner continuum. Our results presented in Fig.~\ref{fig:udep}(e) indicate that the same behavior can be observed in the case of degenerated bands at $U\gg t$. It is worth noting that the overall energy scale of the spin excitation in this phase does not depend strongly on the values of the Hubbard or Hund interaction, but rather on the kinetic energy of the electrons (controlled by, i.e., by the density of the electrons $n$).

\subsubsection{Hund interaction dependence}
\label{sec:hund}

\begin{figure}[!htb]
\includegraphics[width=1.04\columnwidth]{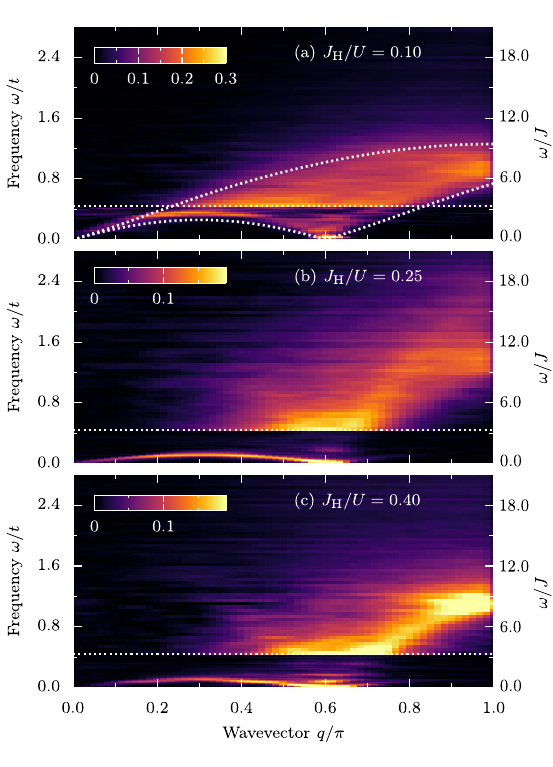}
\caption{Hund interaction $J_\mathrm{H}/U$ dependence of the dynamical spin structure factor $S(q,\omega)$ for finite doping $x=0.4$. Panels (a-c) depict results for $J_\mathrm{H}/U=0.10,0.25,0.40$, respectively. The results below (above) the white dashed line are evaluated with frequency resolution $\Delta\omega/t=0.04$ ($\Delta\omega/t=0.004$).}
\label{fig:hund}
\end{figure}

Since the overall energy scale of the spin excitations in orbital-RVB phase is given by $J=2t^2/(U+J_\mathrm{H})$, the value of the Hund interaction ($J_\mathrm{H}/U$ ratio in our notation) can also influence the overall behavior. For example, one can expect that for a large Hubbard interaction $U$ but vanishing Hund exchange $J_\mathrm{H}\to0$ the spin excitations $S(q,\omega)$ should be akin to a single-band $t$-$J$ result (i.e., two-spinon continuum). Our results for $J_\mathrm{H}/U=0.1$, presented in Fig.~\ref{fig:hund}(a), indeed confirm such scenario. The latter is also expected from the analysis of the $J_\mathrm{H}$-$U$ phase diagram \cite{Jazdzewska2023}. I.e., for $x=0$, the transition between single-band $S=1/2$ and two-orbital $S=1$ physics is given by $J_\mathrm{H}=t^2/U$. Since for finite doping $x\ne0$ the transition occurs \cite{Mierzejewski2024} at a larger value of interaction $U$, and can expect that a similar relation $J_\mathrm{H}\propto b\,t^2/U$, with $b>1$.

For sufficiently large values of the Hund interaction, the system enters the orbital-RVB phase. Our results indicate that within this phase, the results don't depend strongly on the specific value of $J_\mathrm{H}/U$. Such a behaviour is presented in Fig.~\ref{fig:hund}(b,c), where we present $S(q,\omega)$ for $J_\mathrm{H}/U=0.25$ and $0.4$ respectively. Our results indicate that the low-frequency (magnon-like) part is just renormalized by the change in the effective spin exchange value $J=J(J_\mathrm{H})$, as expected. Similarly, the high-frequency incoherent weight at $q>2k_\mathrm{F}$ sharpens, but overall "continuum" is still clearly visible.

It is important to note that for sufficiently large of Hubbard interaction $U$ (and consequently for large $J_H$) the system will enter the ferromagnetic phase \cite{Moreo2025}. For the values considered in this work (e.g., for $U/W=3$ presented in Fig.~\ref{fig:hund}) the system remains in the orbital-RVB phase up to maximal possible $J_\mathrm{H}/U=0.4$ (for $J_\mathrm{H}/U>0.4$ the interorbital interaction $U^\prime$ takes a nonphysical negative value). However, for elevated values of interaction $U\gg t$, we expect that the orbital-RVB phase is stable only in a narrow (non-zero) range of $J_\mathrm{H}/U$ ratio, giving way to the ferromagnetic phase for larger $J_\mathrm{H}$.

\section{Conclusions}
\label{sec:conc}

We have analyzed the ground state and the static $S(q)$ and dynamic $S(q,\omega)$ spin structure factor, i.e., the spin ordering and spin excitations, in the hole-doped $S=1$ Heisenberg model. The latter have to be investigated using the two-orbital Hubbard-Kanamori Hamiltonian, rather than the $S=1/2$ case, where the single-orbital Hubbard model and the $t$-$J$ model capture the overall physics. Primarily, we have focused on the so-called orbital-RVB phase, i.e., the valence-bond-like ground state at finite doping with properties inherited from the AKLT state of the $S=1$ Heisenberg model.

Our results clearly show that finite doping $x\ne0$ of the two-orbital HK model leads to a few distinctive magnetic phases. (i) At $U\,,J_\mathrm{H}\gg t$, a gapless ferromagnetic phase is stabilized due to double-exchange mechanism \cite{Moreo2025}. (ii) At small values of interaction (below the orbital-RVB phase), the maximum of the static structure factor (which in 1D determines the short-range magnetic order) follows the Fermi vector, and the spin excitations resemble those of the single-orbital Hubbard model at finite interaction $U$ (i.e., gapless two-spinon continuum). In the most challenging region $U\sim{\cal O}(W)$, i.e., when the orbital-RVB state is stabilized, our results indicate the presence of two magnetic phases. (iii) For $U/W\lesssim 3$ within orbital-RVB, when $q_\mathrm{max}\simeq 2k_\mathrm{F}$, the spin spectrum exhibits a gapped, "coherent" magnon with vanishing intensity in the $q\to0$ limit. These features are akin to the spectrum of the undoped (half-filled) case. (iv) On the other hand, close to the transition to the ferromagnetic phase (for $U/W\geq 3$), the system is in the gapped spiral phase with the excitations similar to the $J_1$-$J_2$ Heisenberg model.

Although experimentally controlled carrier doping in $S=1$ Haldane chains is challenging, several material platforms demonstrate that this regime is physically meaningful and experimentally accessible \cite{Malvezzi2001,Zaliznyak2001,Kenzelmann2002}. In particular, hole-doped nickelate chains such as Y$_{2-x}$Sr$_x$BaNiO$_5$ have long been known to realize doped $S=1$ backgrounds; photoemission spectroscopy shows a significant increase in spectral weight at the top of the valence band upon doping \cite{Fagot2003}. Early theoretical work \cite{Dagotto1996} on the nickelate chain compound Y$_2$BaNiO$_5$ demonstrated that a multiband model including oxygen and nickel orbitals gives rise to mobile holes interacting with a $S=1$ spin chain, producing sub-gap states in the dynamical spin structure factor consistent with experimental observations. Furthermore, magnetization and susceptibility studies \cite{Janod2001} on Ca-doped Y$_{2-x}$Ca$_x$BaNiO$_5$ reveal a crossover to a spin-glass–like state at low temperatures, confirming that doping strongly modifies magnetic correlations in the $S=1$ chain background. Consequently, while fully tunable doping in ideal $S=1$ chains remains experimentally rare, existing solid-state nickelate compounds already enter regimes where predictions such as doping-dependent shifts of magnon spectral weight and incommensurate magnetic correlations should be testable. Beyond solid-state materials, cold-atom quantum simulators of multiorbital Hubbard systems have recently achieved the Haldane phase \cite{Sompet2022,Richaud2022} and allow tunable doping through particle imbalance. Thus, while fully tunable doping in ideal $S=1$ chains is experimentally rare, several platforms already enter regimes where the physics uncovered here - such as the persistence of a gapped magnon branch, the doping-dependent shift of the spectral maximum to $q=2k_\mathrm{F}$, and the emergence of broad high-energy spectral weight - should be detectable.

The behavior we uncover also has potential implications for higher-dimensional doped antiferromagnets, particularly multiorbital systems where $S>1/2$ local moments arise from Hund’s coupling \cite{Glasbrenner2015,Wang2015,Georges2024}. A central question in two-dimensional correlated materials - including iron-based superconductors and doped nickelates - is how itinerant carriers propagate in an antiferromagnetic background and whether coherent magnon excitations survive in the presence of doping. The latter question are encapsulated in the numerus studies of doped (two-dimensional) spin liquids \cite{Baskaran1987,Anderson1987-2}. Our results demonstrate that (at least in 1D), "coherent" long-wavelength magnons and "incoherent" spinons coexist within an orbital-RVB phase. This coexistence mirrors aspects of the “Hund-metal” phenomenology \cite{Georges2013} in higher dimensions, where local-moment physics and itinerancy intertwine.

\begin{acknowledgments}
J.H. acknowledges grant support by the National Science Centre (NCN), Poland, via Sonata BIS project no. 2023/50/E/ST3/00033. The calculations have been carried out using resources provided by the Wroclaw Centre for Networking and Supercomputing (\url{http://wcss.pl}).\\
The data that support the findings of this article are openly available \cite{opendata}.
\end{acknowledgments}

\bibliography{dopedhk.bib}

\end{document}